\documentclass[noshowpacs,prl,aps,superscriptaddress,floatfix, twocolumn,footinbib,longbibliography]{revtex4-2}
\usepackage[utf8]{inputenc}
\usepackage{cuted}
\usepackage{lineno,mathtools,url}
\usepackage{amsmath}
\usepackage{soul}
\usepackage{amssymb,mathrsfs}
\usepackage{physics}
\usepackage{orcidlink}

\usepackage{graphicx}
\usepackage{cuted}
\usepackage{hyperref}

\usepackage{comment}
\usepackage{dsfont}
\usepackage[dvipsnames]{xcolor}
\hypersetup{
     colorlinks=true,
     linkcolor=blue,
     filecolor=blue,
     citecolor = blue,
     urlcolor=blue,
   }

\newcommand{\ad}{{\hat{a}^\dagger}}
\renewcommand{\a}{\hat{a}}
\newcommand{\bd}{{\hat{b}^\dagger}}
\renewcommand{\b}{\hat{b}}
\newcommand{\n}{\hat{n}}
\newcommand{\I}{\mathrm{I}}
\newcommand{\B}{\mathrm{B}}
\newcommand{\IB}{\mathrm{IB}}
\newcommand{\tc}{t_\mathrm{c}}
\newcommand{\Ep}{E_\mathrm{p}}
\newcommand{\Ec}{\Delta E_\mathrm{c}}
\newcommand{\G}{{G}^{(2)}_{\mathrm{IB}}}

\begin{document}

\title{
Scale invariance of the polaron energy at the Mott--superfluid critical point }

\author{M. Čufar\orcidlink{0000-0003-0734-2719}}
\affiliation{Te Whai Ao --- Dodd-Walls Centre for Photonic and Quantum
 Technologies, Auckland 0632, New Zealand}
\affiliation{Centre for Theoretical Chemistry and
Physics, New Zealand Institute for Advanced Study,
Massey University, Private Bag 102904, North Shore,
Auckland 0745, New Zealand}

\author{R. Alhyder\orcidlink{0000-0002-0853-0714}}
\affiliation{Institute of Science and Technology Austria
(ISTA), Am Campus 1, 3400 Klosterneuburg, Austria
}

\author{C. J. Bradly\orcidlink{0000-0002-5413-777X}}
\affiliation{Department of Mathematics, Swinburne University of Technology, Hawthorn, VIC 3122, Australia}

\author{V. E. Colussi\orcidlink{0000-0002-0972-6276}}
\affiliation{Pitaevskii BEC Center, CNR-INO and Dipartimento di Fisica, Università di Trento, I-38123 Trento, Italy}

\author{G. M. Bruun\orcidlink{0000-0003-4764-1668}}
\affiliation{Center for Complex Quantum Systems, Department of Physics and Astronomy, Aarhus University, Ny Munkegade 120, DK-8000 Aarhus C, Denmark
}%

\author{J. Brand\orcidlink{0000-0001-7773-6292}}
\affiliation{Te Whai Ao --- Dodd-Walls Centre for Photonic and Quantum
Technologies, Auckland 0632, New Zealand}
\affiliation{Centre for Theoretical Chemistry and
Physics, New Zealand Institute for Advanced Study,
Massey University, Private Bag 102904, North Shore,
Auckland 0745, New Zealand}

\author{A. Recati\orcidlink{0000-0002-8682-2034}}
\affiliation{Pitaevskii BEC Center, CNR-INO and Dipartimento di Fisica, Università di Trento, I-38123 Trento, Italy}
\affiliation{Trento Institute for Fundamental Physics and Applications, INFN, 38123 Trento, Italy}

\begin{abstract}
Continuous quantum phase transitions are characterized by an order parameter and correlation functions that are often challenging to access experimentally or in direct numerical simulations. The energy of an added impurity can 
on the other hand be probed by established polaron spectroscopy, or numerically with Monte Carlo methods.
We provide evidence from ground-state quantum Monte Carlo calculations that the energy of a  mobile impurity interacting weakly with a surrounding lattice Bose gas provides access to the critical behavior of the Mott insulator--superfluid phase transition. Finite-size scaling of the energy reveals that its value is scale invariant at the critical point of the quantum phase transition, and we extract
a scaling exponent that is currently unexplained by theory. For a small lattice we further observe a flattening of the impurity-boson density-density correlations at the critical point, which hints at a divergence of a corresponding length scale in the thermodynamic limit. 
Our results suggest that impurity spectroscopy represents a useful way to probe the critical properties of quantum phase transitions in general.
\end{abstract}

\maketitle

\emph{Introduction ---} Quantum phase transitions are a hallmark of many-body physics, where quantum fluctuations drive the system’s ground state across a quantum critical point~\cite{sachdev2011}. 
At the critical point, long-range correlations arise and the system exhibits universal scaling behavior, i.e., various observables follow a power-law independent of the microscopic details of the system~\cite{fisher1972}. These transitions are typically characterized by an order parameter, an expectation value of an operator that vanishes in one phase and takes a non-zero value in the other. However, identifying the appropriate order parameter for a given phase transition can be nontrivial, and even when known, it is often difficult to access experimentally. Probing such quantities typically requires high-resolution, phase-sensitive measurements that are challenging to realize, leaving many quantum transitions only partially mapped~\cite{cardy2012finite,Matthias_Vojta_2003}.

Recently, it has been argued that the energy of a mobile impurity embedded in a many-body system can serve as a sensitive probe of quantum phases~\cite{CamachoGuardian2019,alhyderMobileImpurityProbing2022a,colussiLatticePolaronsSuperfluid2023a,vashishtChiralPolaronFormation2025,alhyder2025,massignanPolaronsAtomicGases2025,hartweg2025,grusdtImpuritiesPolaronsBosonic2025,dominguez-castroPolaronsBipolaronsRydbergdressed2026}. Through interactions with the host medium, the impurity becomes dressed by excitations forming a quasiparticle called a polaron, whose properties reflect the surrounding environment. In the regime where the impurity–host interaction is weak compared to the characteristic interaction scale of the host system, the polaron behaves as a passive probe, with its energy reflecting properties of the surrounding medium without significantly disturbing it~\cite{colussiLatticePolaronsSuperfluid2023a,alhyder2025}. Extracting the polaron energy with high-fidelity  techniques such as radio-frequency, Ramsey, or modulation spectroscopy~\cite{skouNonequilibriumQuantumDynamics2021,etrych2024universal,Henke2025,Alhyder2026}, may thus provide insights into the properties of the system across a quantum phase transition. As such, it offers a powerful alternative to conventional probes that often rely on bulk correlation functions, which are significantly more challenging to measure experimentally.

A particularly relevant setting is the Bose–Hubbard model of bosons in a tight-binding lattice with repulsive interactions.
As the ratio of hopping to interaction strength is increased, the system undergoes a continuous quantum phase transition from a gapped Mott insulator to a gapless superfluid at fixed integer density~\cite{fisher_boson_1989, sachdev2011}, as realized experimentally with ultracold atoms in optical lattices~\cite{greiner2002quantum,Jordens2008,Sherson2010,Su2025}. The critical behavior of the model has been investigated using quantum Monte Carlo simulations~\cite{krauthMottSuperfluidTransitions1991,capelloSuperfluidMottInsulatorTransition2007,capogrosso-sansoneMonteCarloStudy2008,pollet2012,gazit2013}, non-perturbative renormalization group methods~\cite{rancon_nonperturbative_2011,ranconQuantumXYCriticality2013,ranconHiggsAmplitudeMode2014a,roseHiggsAmplitudeMode2015}, and other analytical approaches~\cite{sengupta2005, podolsky_spectral_2012,hinrichsPerturbativeCalculationCritical2013a, lackiLocatingQuantumCritical2016a,sandersQuantumCriticalProperties2019}.

The Mott-superfluid transition in the two dimensional Bose-Hubbard model is a continuous phase transition in the same universality class as the three-dimensional classical $O(2)$ model~\cite{fisher_boson_1989}.
Here, we use finite-size scaling of Monte Carlo data to show that the addition of a polaron maintains much of the critical properties of the host system with some key differences. We find that the energy of the polaron is itself scale invariant and can serve as a predictor of the location of the critical point.
Furthermore, we show that there is an intimate connection between the polaron energy and the impurity-boson density-density correlation function, which in turn probes the dynamic structure factor of the host system for weakly-coupled impurities. We find that even for small system sizes the impurity-boson correlations flatten at the transition point, which provides a further observable characteristic of the quantum phase transition accessible to quantum gas microscopy experiments~\cite{Bakr}.  
However, despite the intimate connection between the polaron and the host system, we also find that the crossover scaling exponent for the polaron energy with value $\approx0.74\pm0.26$ is distinctly different from the crossover exponent associated with the host system  $1/\nu\approx1.49$~\cite{sachdev2011}, which controls the divergence of the correlation length, or equivalently the spectral gap across the phase transition. The origin of this critical exponent represents an intriguing theoretical challenge.

\emph{The model ---} The Bose-Hubbard Hamiltonian describing the host system is
\begin{equation}\label{eq:bath}
  \hat{H}_\B = -t\sum_{\langle \vb{i},\vb{j}\rangle} \ad_{\vb{i}}\a_{\vb{j}} + \frac{U}{2}\sum_{\vb{i}}\n_{\vb{i},\B} (\n_{\vb{i},\B} - 1)\,,
\end{equation}
where $\ad_{\vb{i}}$ ($\a_{\vb{i}}$) creates (destroys) a boson at site $\vb{i}$, $\langle\vb{i},\vb{j}\rangle$ goes over all neighboring sites, $\n_{\vb{i},\B}=\ad_{\vb{i}}\a_{\vb{i}}$ is the number operator, $t$ the hopping parameter and $U>0$ the repulsive boson-boson interaction strength. We consider the model on a two-dimensional $L\times L$ square lattice with periodic boundaries filled with $N_\B$ bosons, focusing on the case of unit filling where $n_\B\equiv N_\B/L^2 =1$. At fixed $U$, this model undergoes a quantum phase transition as the hopping strength $t/U$ crosses a critical value, $\tc/U$.
We add a single, interacting impurity to the model by modifying the Hamiltonian as
\begin{equation}\label{eq:hamiltonian}
  \hat{H} = \hat{H}_\B + \hat{H}_\mathrm{I}  + U_{\IB}\sum_{\vb{i}}\n_{\vb{i},\I}\n_{\vb{i},\B}\,,
\end{equation}
where $\hat{H}_\mathrm{I} = -t\sum_{\langle \vb{i},\vb{j}\rangle} \bd_{\vb{i}}\b_{\vb{j}}$ is the impurity Hamiltonian, $U_\IB$ is the impurity-boson interaction strength, $\bd_{\vb{i}}$ ($\b_{\vb{i}}$) is the creation (destruction) operator for the impurity, and $\n_{\vb{i},\I} = \bd_{\vb{i}}\b_{\vb{i}}$.

\emph{Computational methods ---} To compute the ground state energy of $\hat{H}$, we employ full configuration interaction quantum Monte Carlo (QMC)~\cite{boothFermionMonteCarlo2009,Cufar2026}, which is a projector Monte Carlo method.
The algorithm iterates from an arbitrary starting vector $\vb{c}^{(0)}$ according to the rule
\begin{equation}\label{eq:fciqmc}
  \vb{c}^{(n+1)} = \left[\vb{1} + \mathrm{d}\tau \, \left(S^{(n)} - \vb{H} \right)\right]\vb{c}^{(n)}\,,
\end{equation}
where $\vb{1}$ is the identity matrix, $\mathrm{d}\tau$ a (small) time step, $\vb{H}$ is a matrix representation of $\hat{H}$ and $S^{(n)}$ is a scalar energy shift, which is updated after every step in a way that keeps the 1-norm of $\vb{c}^{(n)}$ approximately constant~\cite{boothFermionMonteCarlo2009,yangImprovedWalkerPopulation2020}. As long as the time step is small enough, Eq.~\eqref{eq:fciqmc} has a stable fixed point where $c^{(\infty)}$ is the ground state eigenvector and $S^{(\infty)}$ the ground state energy of $\vb{H}$.

The scheme is applied stochastically such that Eq.~\eqref{eq:fciqmc} is fulfilled in expectation but only a small number of nonzero vector elements need to be stored and matrix elements are evaluated on the fly.
Equation~\eqref{eq:fciqmc} is applied for many steps and, after an equilibration phase, the values of $S^{(n)}$ fluctuate around $E$, the exact ground state energy of $\hat{H}$. Averaging this data gives an accurate estimate of the energy $E$, albeit with a controllable and usually small population control bias, which is caused by the noise in the algorithm~\cite{brandStochasticDifferentialEquation2022,ghanemPopulationControlBias2021}. 
To control this bias, we use importance sampling, which modifies $\vb{H}$ by similarity transform in a way that reduces noise in the algorithm.
This allows us to accurately sample ground states of systems with Hilbert space dimension up to $\approx10^{60}$.

\emph{Polaron energy as a scaling function ---} The polaron energy $\Ep$ is given by the difference between the ground state energy $E$ of the system with ($U_\IB\neq 0$) and without ($U_\IB=0$) the impurity:
\begin{equation}\label{eq:polaron-energy}
  \Ep = E(U_\IB) - E(0)\,.
\end{equation}
\begin{figure*}
\includegraphics{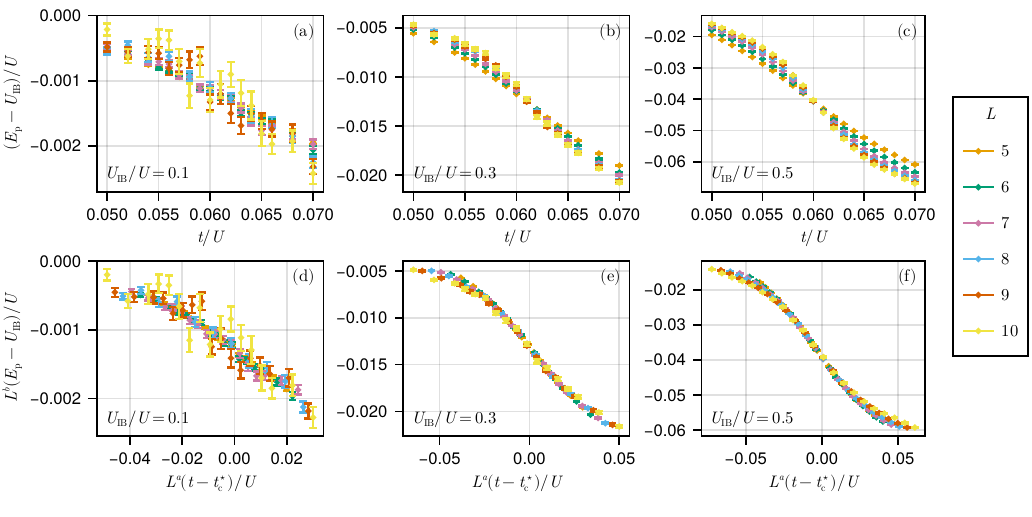}
\caption{\label{fig:big-plot}Finite size scaling of the polaron energy for a Bose-Hubbard lattice at unit filling with an  impurity-boson interaction strength $U_\IB$ as labeled. The top row (a--c) shows the polaron energy $\Ep$ as a function of $t/U$. An apparent crossing of the curves for different system size $L$ (as per legend) near critical value $\tc/U \approx 0.06$ of the Mott insulator--superfluid phase transition indicates the scale invariance of $\Ep$. The bottom row (d--f) shows the same data scaled according to the ansatz from Eq.~\eqref{eq:ansatz} with optimized parameters. The parameter values are shown in Fig.~\ref{fig:parameters}. 1$\sigma$ error bars are used throughout.}
\end{figure*}
\begin{figure}
  \centering
  \includegraphics{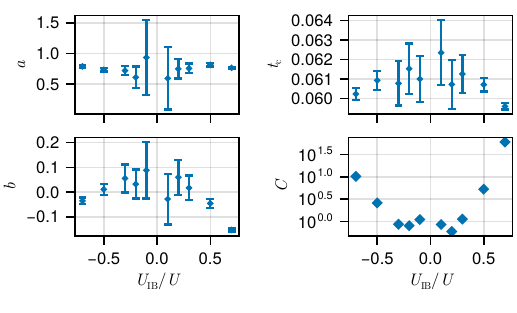}\caption{\label{fig:parameters}Scaling parameters. Shown are the fitted parameters $a$, $b$, and $\tc$ of the scaling ansatz of Eq.~\eqref{eq:ansatz} vs.~the interaction strength $U_\IB$. The data was collected for system sizes $L$ between 5 and 10 in a narrow window close to the critical interaction strength $\tc/U$. Values of the cost function $C$ close to 1 indicate that the scaling ansatz is well supported by the data.}
\end{figure}
The polaron energy as a function of the hopping strength $t/U$ for various system sizes $L$ is shown in the top row of Fig.~\ref{fig:big-plot}, where the different panels~(a--c) correspond to different values of the coupling strength $U_{\IB}$. The curves for different values of $L$ are found to cross near the critical hopping strength for the quantum phase transitions $\tc/U \approx 0.06$~\cite{krauthMottSuperfluidTransitions1991,capelloSuperfluidMottInsulatorTransition2007,capogrosso-sansoneMonteCarloStudy2008}, which indicates the polaron energy is scale invariant at the transition point. It further suggests that these curves may be connected by finite size scaling~\cite{cardy2012finite}.
To investigate this we make a scaling hypothesis using the standard ansatz
\begin{equation}
  \label{eq:ansatz}
  \frac{E_\mathrm{p} - U_\IB}{U} = L^{-b} f\left(L^a \frac{t - \tc}{U}\right)\,,
\end{equation}
where the parameters we want to estimate are the crossover exponent $a$,  the leading exponent $b$, and the location of the critical point $\tc$, with $f$ an unknown analytic function. Because a single impurity is only a local effect in the thermodynamic limit, $f$ only has explicit dependence on $L$ and $t/U$ while $U_\IB$ is only expected to have an implicit effect on the non-universal behavior of $f$. 
When the parameters of Eq.~\eqref{eq:ansatz} are chosen correctly, and the data is plotted as $L^b (\Ep - U_\IB)/U$ versus $L^a (t - \tc)/U$, the data near the transition point should collapse onto a single curve~\cite{sachdev2011}. To find these parameters, we define a cost function $C(a,b,\tc)$ (see End Matter), that measures how collapsed the data is. In a modification of the procedure of Ref.~\cite{houdayerLowtemperatureBehaviorTwodimensional2004} we interpolate the data across $t/U$ for each value of $L$ and measure the distance each such curve is from an average, in units of the QMC error bars. This cost function is similar to the reduced $\chi^2$ measure of goodness of fit, and as such, values close to 1 indicate the data collapses well.

The rescaled data is shown in the bottom row of panels~(d--f).
We see that the data collapses well in all three cases, however, the data with $U_\IB/U=0.1$ is rather noisy, and at $U_\IB/U=0.5$, the collapse begins to show signs of breaking down. This can be made clearer by looking at Fig.~\ref{fig:parameters}, where the values of the fitted parameters and $C$ are plotted across a range of $U_\IB/U$. We see that $C$ increases as $|U_\IB/U|$ becomes stronger, which indicates the collapse only works well when interactions are weak. This is expected behavior for finite systems with fixed particle number --- when interactions are strong, the impurity has a significant effect on the underlying system and can no longer be seen as a passive probe~\cite{alhyder2025}.
The weak $U_\IB/U$ data has relatively more noise because the QMC error bar primarily depends on the system size, while the span of $\Ep/U$ values depends strongly on $U_\IB/U$. This is reflected in the size of the error bars on the parameter estimates in Fig.~\ref{fig:parameters}.
The data provides strong support for the scaling hypothesis being valid in the regime of weak impurity-boson interaction  $|U_\IB/U| \lesssim 0.5$. We find the critical point to occur at $\tc/U = 0.061 \pm 0.0013$, which is consistent with other values in the literature~\cite{krauthMottSuperfluidTransitions1991,capelloSuperfluidMottInsulatorTransition2007,capogrosso-sansoneMonteCarloStudy2008}. In addition, we estimate $b = 0.05 \pm 0.078$, which is consistent with zero, making $\Ep$  scale invariant at the critical point. Finally, our estimate for the crossover exponent is $a = 0.74 \pm 0.26$. 
Finite size effects may cause a bias that we estimate to be less than 10\% by calculating the crossover exponent of the host system's charge gap, which is accurately known in the literature~\cite{sachdev2011} (see End Matter and Fig.~\ref{fig:chargegap}).

\emph{Impurity-boson density-density correlations ---}
To further examine the connections between the 
polaron energy and the critical properties 
of the host system, we now consider the
impurity-boson density-density correlation function
\begin{equation}\label{eq:g2}
  \G(\vb{d}) = \sum_{\vb{i}} \langle\n_{\vb{i},\I}\n_{\vb{i}+\vb{d},\B}\rangle\,,
\end{equation}
An application of the Hellman-Feynman theorem to the polaron energy in Eq.~\eqref{eq:polaron-energy} yields
\begin{align}\label{eq:hellmann}
  \frac{d \Ep}{dU_\IB} & =  \sum_{\vb{j}} \langle\n_{\vb{j},\I}\n_{\vb{j},\B} \rangle \equiv
   \G(\vb{0})\,,
\end{align}
which shows that $\Ep$ directly probes the local impurity-boson correlations $\G(\vb{0})$. This relation is exact for arbitrary interaction strengths. 
For weak impurity-boson 
$U_\IB$, the polaron energy can further be related to the properties of the host system by a perturbative expansion in $U_\IB$~\cite{bigue2022,alhyder2024} (see End Matter for details)
\begin{align} \label{eq:EpSkomega}
  \Ep &= U_\IB n_\B +  U_\IB^2  \sum_{\vb{k}} \int d\omega \frac{S(\vb{k},\omega)}{\epsilon_{\vb{0}}  - \epsilon_{\vb{k}} - \omega}
+ O(U_\IB^3),
\end{align}
where $S(\vb{k},\omega)$ is the dynamic structure factor encapsulating the dynamic density-density correlations of the bosons. 
Equation~\eqref{eq:EpSkomega} demonstrates how the polaron energy directly relates to non-local correlations in the unperturbed host system. This motivates  further examination of $\G$ across the Mott insulator--superfluid transition. Note that a successful theoretical investigation should include terms involving more than one particle-hole excitation, since in the single particle-hole channel the polaron energy vanishes at the quantum critical point~\cite{Punk2013,alhyder2025}.

\begin{figure}
    \centering
    \includegraphics{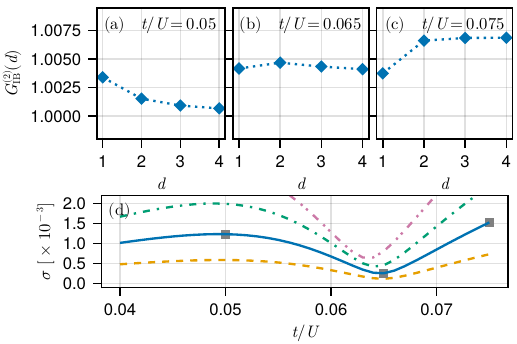}
    \caption{\label{fig:g2-1d}
    Flattening of non-local impurity-boson correlations in a small system with $L=4$. Panels~(a--c) show the impurity-boson correlation function  $\G(d)$ as a function of the rectilinear distance ($L^1$ metric) $d$ of a host boson from the position of the impurity at an interaction strength of $U_\IB/U=0.2$. Panel~(d) shows the standard deviation $\sigma$ (see Eq.~\eqref{eq:sigma}) of $\G(d)$ for $d>0$ vs.~$t/U$, with gray squares marking the values corresponding to panels~(a--c). The lines from bottom to top correspond to $U_\IB/U=0.1,0.2,0.3,0.4$. A distinct minimum of $\sigma$, indicating a flattening of correlation function, is observed near the expected critical value of $t/U$ for the Mott-insulator -- superfluid quantum phase transition.}
\end{figure}


\emph{Small system behavior ---}
Encouraged by the numerically observed scale invariance of $\Ep$, we examine the impurity-boson correlations for a small system with exact diagonalization calculations.
Figure~\ref{fig:g2-1d}~(a--c) shows the correlation function $\G(d)$ for $d\equiv \Vert\vb{d}\Vert_1>0$ for a small lattice with $L=4$. The values of $\G(0)$ are not shown and are presented in End Matter.
At small $t/U$ in Fig.~\ref{fig:g2-1d}~(a) the host boson density is increased in the immediate vicinity ($d=1$) of the impurity compared to the asymptotic value at larger distances, as expected for an insulating phase~\cite{Kobayashi1996,Shiino1998}.
Conversely, at large $t/U$ in Fig.~\ref{fig:g2-1d}~(c), the host boson density is depressed near the repulsive impurity before healing to an asymptotic value at larger distances \footnote{Note that the asymptote being larger than 1 is a result of the small size of the system combined with the strong suppression of the host boson density at the same site as the impurity ($d=0$, not shown).}. This behavior is reminiscent of the weakly-interacting (superfluid) Bose gas~\cite{Pitaevskii2016a}.
Between these two regimes, at the intermediate value $t/U=0.065$ as shown in panel~(b), the correlations are almost entirely flat. 
We quantify the flatness through the standard deviation $\sigma$ of the non-local impurity-boson correlation function, namely
\begin{equation}\label{eq:sigma}
    \sigma^2 =  \frac{1}{L-1}\sum_{d > 0}{\left[\G(d) - \bar{g}\right]}^2
\end{equation}
where $\bar{g} = \frac{1}{L} \sum_{d > 0} \G(d)$ is the non-local average. The standard deviation $\sigma$ is shown in Fig.~\ref{fig:g2-1d}~(d) as a function of the hopping strength $t/U$ and for several values of $U_\IB/U$.
There is a distinct minimum in $\sigma$ near $\tc/U \approx 0.065$ suggesting that the flatness of the impurity-boson correlations are another indicator of the transition for weak $U_\IB/U$. 
This flatness can be rationalized as a finite size precursor of a divergent correlation length in the thermodynamic limit, and is consistent with the observed scale invariance of the polaron energy near the critical point. 
The location of the minimum of $\sigma$ in Fig.~\ref{fig:g2-1d}~(d) is seen to move towards smaller values of $t/U$ for larger impurity strengths $U_\IB$, which we attribute to non-perturbative effects.

In a smaller system with $L=3$ (data not shown) we find similar behavior with a local minimum of $\sigma$ at a nearly unchanged location of $t/U\approx0.066$. Based on these observations, we hypothesize that the scale invariance observed for the polaron energy (which measures local impurity-boson correlations) also extends to the non-local impurity-boson correlations.
As the impurity-boson density-density correlation function $\G(\vb{d})$ can directly be measured in quantum gas microscope experiments~\cite{Bakr}, observing the flattening of the non-local correlations provides an accessible indicator for the quantum phase transition precursor in even relatively small finite systems.

\emph{Outlook and conclusion ---} 
Our numerical simulations indicate that the energy of a single impurity immersed in a two-dimensional Bose Hubbard model follows a scaling law when the host system undergoes a Mott insulator--superfluid phase transition. Specifically, our data supports the hypothesis that the polaron energy is scale invariant at the transition point. 
Our estimate for the crossover exponent of the polaron energy $a=0.74\pm0.26$ is distinctly different from
the known critical properties of the host system~\cite{sachdev2011}.
While we have theoretically derived a direct connection between the polaron energy of a weakly-coupled impurity and
the density-density correlations of the host system,
a microscopic explanation of our results is an interesting and open problem to the best of our knowledge. 
From a broader perspective, our results open up interesting research directions concerning how impurities can provide powerful new ways to probe the critical properties of quantum phase transitions, which  
are hard to measure by other means. This is particularly relevant for high precision quantum simulation experiments with cold atoms where the interaction, number of particles and lattice sites can be controlled~\cite{Koehn2025,Leonard2023}.

\begin{acknowledgments}
\emph{Acknowledgements ---} We thank Nicolas Dupuis and Adam Rançon for fruitful discussions.
M.Č. and J.B. received funding from the Marsden Fund of New Zealand (Contract No.\ MAU 2007), from government funding managed by the Royal Society of New Zealand Te Apārangi, and acknowledge support by the New Zealand eScience Infrastructure (NeSI) high-performance computing facilities in the form of a merit project allocation. This research used resources of the Oak Ridge Leadership Computing Facility at the Oak Ridge National Laboratory, which is supported by the Office of Science of the U.S.\ Department of Energy under Contract No.\ DE-AC05-00OR22725.
R.A. acknowledges support from the Austrian Academy of Science ÖAW grant No.\ PR1029OEAW03. A.R. received financial support from Provincia Autonoma di Trento and and from the Italian National Institute for Nuclear Physcis (INFN) through the RELAQS project. G.M.B. received supported from the Danish National Research Foundation through the Center of Excellence ``CCQ'' (Grant agreement No.\ DNRF156), the Independent Research Fund Denmark- Natural Sciences via Grant No.\ DFF -8021-00233B.  This research was supported in part by the National Science Foundation under Grant No.\ NSF PHY-1748958.
\end{acknowledgments}
\bibliography{biblio.bib}


\newpage

\appendix

\section{End Matter}

\emph{The charge gap ---}\label{app:chargegap}
An important quantity characterizing the Mott insulator-superfluid phase transition is the charge gap, which captures the asymmetry of changes in the ground state energy due to either adding or removing a single particle. It is defined as
\begin{equation}
    \Ec = \frac{1}{2}\left[E(L^2+1) + E(L^2-1)\right] - E(L^2)\,,
\end{equation}
where $E(N)$ is the ground state of a system with $N$ particles. In the thermodynamic limit, $\Ec$ vanishes for a superfluid, and a positive nonzero value characterizes the Mott-insulating regime.

\begin{figure}
    \centering
    \includegraphics{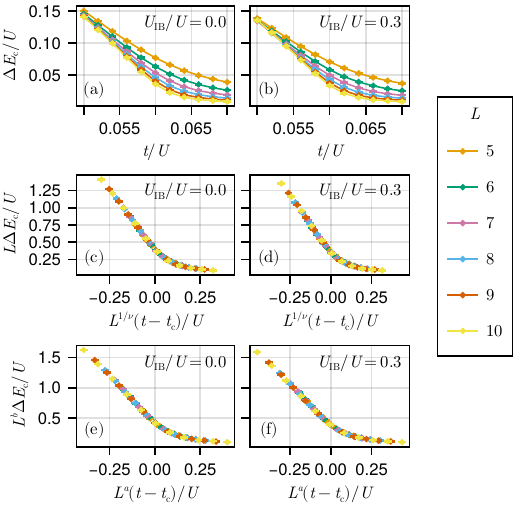}
    \caption{\label{fig:chargegap}The charge gap $\Ec/U$ as a function of $t/U$ for the Bose-Hubbard lattice and unit filling without impurity (a, c, e) and with repulsive impurity (b, d, e). The top panels~(a,b) show the unmodified data. The middle panels~(c,d) show the data scaled with known parameters $a=1/\nu\approx 1.49$, $b=1$, while the lower panels~(e,f) show the data scaled with fitted parameters $a\approx1.598\pm0.009$, $b\approx1.06\pm0.02$ (e) and $a\approx1.646\pm0.008$, $b\approx1.07\pm0.01$ (f).}
\end{figure}

In Fig.~\ref{fig:chargegap}~(a) we show the raw charge gap data for the Bose-Hubbard host without impurity for different system sizes. In panel~(c), we show the same data scaled using the equivalent of Eq.~\eqref{eq:ansatz} with the parameters known from the Ref.~\cite{podolsky_spectral_2012}, $a=1/\nu=1.49$ and $b=1$, while in Fig.~\ref{fig:chargegap}~(e), we show it scaled with parameters $a$ and $b$ estimated with our procedure. We see the fitted estimates $a\approx1.598\pm0.009$, $b\approx1.06\pm0.02$ are close to the expected values $a=1/\nu\approx 1.49$, $b=1$.  We attribute the deviation from the literature values to the fact that a very limited range of system sizes $L$ is used in the estimation, which causes a small systematic bias. In Fig.~\ref{fig:chargegap}~(b,d,f), we repeat the procedure with an interacting impurity ($U_\IB/U=0.3$). We notice that the estimated parameters change in the third digit, which is barely significant and larger than the $1\sigma$ statistical error, indicating that the impurity may have a small effect on the underlying phase transition. We expect this effect to disappear in the thermodynamic limit, where a single impurity should not be able to affect the quantum phase transition. We note that the bias in the estimated exponents from finite-size effects is of the order of or less than 10\% in all cases.

\emph{The cost function ---}\label{app:costfun}
We estimate the scaling parameters by optimizing the parameters of a scaling ansatz to minimize a cost function. The cost function we use is based on the one presented in Ref.~\cite{houdayerLowtemperatureBehaviorTwodimensional2004}, however, we have modified it in a way to make it a continuous function of its parameters. This makes the optimization procedure more robust as the discontinuities in the cost function can cause the optimizer to get stuck in local minima.

The cost function is defined as follows. We assume the data consists of $\vb{x},\vb{y},\vb{dy}$ for a variety of system sizes $L$ on the interval $[x_l, x_h]$. The data is already rescaled with some values of the parameters, in this case $x = L^a(t - \tc)/U$ and $y = L^b(\Ep - U_\IB)/U$.
We interpolate the data such that $y_L : [x_l, x_h] \to \mathbb{R}$ and $dy^2_L : [x_l, x_h] \to \mathbb{R}$ are piecewise linear functions, where $y_L$ is an interpolation of the (transformed) data and $dy^2$ is the interpolated square of the error bar. We then define the master curve as a weighted average of the interpolated data, weighted by the error bars
\begin{equation}
  \label{eq:master-curve}
  Y(x) = \frac{\sum_L y_L(x)/dy_L^2(x)}{\sum_{L}1/dy_L^2(x)}\,.
\end{equation}
To compute the cost function, we compute the contribution of each $L$-subset to the cost function. First, define the square deviation of the data and from master curve as
\begin{equation}
  \label{eq:2}
  \delta^2_L(x) = \frac{{\left(y_L(x) - Y(x)\right)}^2}{dy^2_L(x)}\,.
\end{equation}
Then the contribution from $L$ is
\begin{equation}
  \label{eq:integral}
  C_L = \frac{1}{x_h - x_l}\int_{x_l}^{x_h}\delta^2_L(x) dx\,.
\end{equation}
Finally the value of the cost function is given by
\begin{equation}
  \label{eq:4}
  C = \frac{1}{N_L}\sum_L C_L\,,
\end{equation}
where $N_L$ is the number of distinct values of $L$. Like in Ref.~\cite{houdayerLowtemperatureBehaviorTwodimensional2004}, the definition of the cost function is similar to the reduced $\chi^2$ measure. As such, values of $C$ close to 1 indicate the data collapses well.

\emph{Polaron energy and density correlations ---}
%
\begin{figure}
    \centering
    \includegraphics{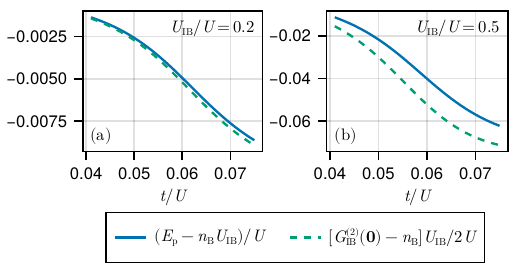}
    \caption{Relationship between the the polaron energy $\Ep$ and the on-site impurity-boson density-density correlations $\G$. As expected from Eq.~\eqref{eq:epsecondorder}, the differences between the two quantities are small for (a) $U_\IB=0.2U$ and become more appreciable for the stronger impurity with (b) $U_\IB=0.5U$. Both computations were done for a system with $L=4$ and $n_\B = 1$.
    }
    \label{fig:g2-of-zero}
\end{figure}
As discussed in the main text, the polaron energy $\Ep$ can be seen as a probe of the on-site impurity-boson density-density correlations $\G(\vb{d})$. This relationship is further evidenced by Fig.~\ref{fig:g2-of-zero}, which shows that $\G(\vb{0})$ and $\Ep$ are closely related. In Eq.~\eqref{eq:hellmann} in the main text, we have shown that the Hellmann Feynman theorem can be used to relate $\Ep$ and $\G(\vb{0})$. We proceed by expanding the impurity-boson correlation function up to first order in $U_\IB$. Using the shorthand notation $\hat{V}= \sum_{\vb{j}} \n_{\vb{j},\I} \n_{\vb{j},\B}$,
\begin{align} \label{eq:Gexpansion}
  &\G(\vb{0}) \equiv \langle \psi| \hat{V} |\psi \rangle \nonumber \\
  &\quad =  \langle \phi_{0,\vb{0}}|  \hat{V}|\phi_{0,\vb{0}}\rangle
  + 2 U_\IB \Re{\langle \phi_{0,\vb{0}}| \hat{V} |  \psi^{(1)} \rangle
  } \nonumber\\
  &\quad + O(U_\IB^2),
\end{align}
to first order in $U_\IB$, where 
\begin{align}
  |\psi\rangle &= |\phi_{0,\vb{0}}\rangle
  + U_\IB |\psi^{(1)}\rangle + O(U_\IB^2) ,
\end{align} 
is the expansion of the full-system ground state, and
\begin{align}
  |\phi_{\nu,\vb{k}}\rangle &= |\phi_\nu\rangle \otimes |\vb{k}\rangle ,
\end{align}
are the eigenstates of the unperturbed Hamiltonian $\hat{H}_\mathrm{B} + \hat{H}_\mathrm{I}$ with energies $E_{\nu, \vb{k}} = E_\nu + \epsilon_{\vb{k}}$. Here, $E_\nu$ are the eigenenergies of the host-system states $|\phi_\nu\rangle$, and the single-impurity plane wave states
$|\vb{k}\rangle = L^{-1}\sum_{\vb{j}}e^{-i\vb{k}\cdot\vb{j}}\,\bd_{\vb{j}}|\mathrm{vac}\rangle$ have the usual lattice dispersion 
\begin{align}
  \epsilon_{\vb{k}} = -2t[\cos(k_x) + \cos(k_y)] . 
\end{align}
The first order correction to the ground state is given by the standard expression
\begin{align}
  |\psi^{(1)}\rangle  &= \sum_{\nu, \vb{k} \ne 0, \vb{0}} \frac{|\phi_{\nu,\vb{k}}\rangle\langle \phi_{\nu,\vb{k}}| \hat{V} |\phi_{0,\vb{0}}\rangle}{E_{0,\vb{0}}- E_{\nu,\vb{k}}} .
\end{align} 
The overlap matrix elements simplify as an integration over the impurity eigenstates and can be carried out explicitly to yield
\begin{align}
  \langle \vb{k}|\hat{V}|\vb{0}\rangle &= 
  \frac{1}{L^2}\sum_{\vb{j}} e^{i\vb{k}\cdot \vb{j}} \n_{\vb{j},\B}
  \equiv \hat{\rho}_{\vb{k}} ,
\end{align}
which is the momentum space particle density operator for the host system. The first order term from Eq.~\eqref{eq:Gexpansion} then simplifies to
\begin{align}
  \langle \phi_{0,\vb{0}}| \hat{V} |  \psi^{(1)} \rangle &= 
  \sum_{\nu, \vb{k} \ne 0, \vb{0}} \frac{\langle \phi_0 |\hat{\rho}_{\vb{k}}^\dagger|\phi_{\nu}\rangle\langle \phi_{\nu}| \hat{\rho}_{\vb{k}} |\phi_{0}\rangle}{E_{0}- E_{\nu} +  \epsilon_{\vb{0}} - \epsilon_{\vb{k}}} \\
  &= \sum_{\vb{k}} \int_{-\infty}^{\infty} d\omega \frac{S(\vb{k},\omega)}{\epsilon_{\vb{0}} - \epsilon_{\vb{k}}- \omega} , 
\end{align}
where
\begin{align}
  S(\vb{k},\omega) &= \sum_{\nu} |\langle \phi_{\nu}| \delta\hat{\rho}_{\vb{k}} |\phi_{0}\rangle|^2 \delta(\omega - E_{\nu}+ E_{0}) ,
\end{align}
is the dynamic structure factor of the host system at zero temperature. The use of the density fluctuation operator $\delta \hat{\rho}_{\vb{k}} = \hat{\rho}_{\vb{k}} - \langle \hat{\rho}_{\vb{k}}\rangle = \hat{\rho}_{\vb{k}} - \delta_{\vb{k},\vb{0}} n_\B$ removes a divergent term and realizes the exclusion of the double ground state, where $\nu=0, \vb{k}= \vb{0}$, from the sum.

Using the fact that the impurity-boson density-density correlation function normalizes to the boson density $n_\B =  N_B/L^2$ in the absence of impurity-boson interactions
\begin{align}
  \G(\vb{0}) |_{U_\IB = 0} &= \langle \phi_{0,\vb{0}}|  \hat{V}|\phi_{0,\vb{0}}\rangle = n_\B ,
\end{align}
we obtain
\begin{align}
  \G(\vb{0}) &= n_\B + 2 U_\IB \sum_{\vb{k}} \int d\omega \frac{S(\vb{k},\omega)}{\epsilon_{\vb{0}}  - \epsilon_{\vb{k}} - \omega} \nonumber \\
  &\quad + O(U_\IB^2) .
\end{align}
Integrating the Hellman-Feynman relation Eq.~\eqref{eq:hellmann} yields Eq.~\eqref{eq:EpSkomega} for
the polaron energy to second order in $U_\IB$ in terms of the properties of the fully interacting host system. Equivalently, this allows us to relate the polaron energy and the local correlations to second order by
\begin{align} \label{eq:epsecondorder}
    \Ep - U_\IB n_\B = \frac{1}{2}U_\IB\left[\G(\vb{0}) - n_\B\right]+ O(U_\IB^3) .
\end{align}
Figure~\ref{fig:g2-of-zero} shows the left hand side and the right hand side of this equation and demonstrates that the remaining third order terms are negligible when the impurity-boson interaction strength is small.
\end{document}